\documentstyle[epsfig]{article}

\begin{document}

\title{
Learning curves for Soft Margin Classifiers
}

\author{
Sebastian Risau-Gusman\dag\S \, and Mirta B. Gordon\ddag\S \\
\dag\ Departamento de F{\'{\i}}sica, Facultad de Ciencias Exactas y Naturales \\ 
Universidad de Buenos Aires, Pabell\'on 1, Ciudad Universitaria \\ 
1428 Buenos Aires, Argentina.\\
\ddag\ Laboratoire Leibniz - IMAG \\
46 ave. Felix Viallet, 38031 Grenoble Cedex, France \\
\S\ D\'epartement de Recherche Fondamentale sur
la Mati\`ere Condens\'ee\\ CEA - Grenoble, 17 rue des Martyrs,
38054 Grenoble Cedex 9, France\\
}

\date{February 28, 2002}
\maketitle

\begin{center}
\begin{abstract}
Typical learning curves for Soft Margin Classifiers (SMCs) 
learning both realizable and unrealizable tasks are determined 
using the tools of Statistical Mechanics. We derive the analytical
behaviour of the learning curves in the regimes of small and large training
sets. The generalization errors present different decay laws towards 
the asymptotic values as a function of the training set size, depending 
on general geometrical characteristics of the rule to be learned. 
Optimal generalization curves are deduced through a fine tuning of the 
hyperparameter controlling the trade-off between the error and the 
regularization terms in the cost function. Even if the task is 
realizable, the optimal performance of the SMC is better than that 
of a hard margin Support Vector Machine (SVM) learning the same rule, 
and is very close to that of the Bayesian classifier.
\end{abstract}

\end{center}


\newpage
\large{

\section{Introduction}

The recently introduced Support Vector Machines~\cite{CoVa} 
(SVM) may be considered as an extension of the perceptron. 
The latter is only able to perform linear separations by a 
hyperplane in input space. When the problem is not linearly 
separable, instead of searching for more complex surfaces in 
input space, SVMs map the input patterns onto a space of 
much higher dimension with the hope that in this 
{\it featurs-space} the task be linearly separable. To cope 
with the problem of the very high dimensionality of the space, 
Cortes and Vapnik~\cite{CoVa}  proposed to find the Maximal 
Stability Perceptron (which is the solution that maximizes 
the distance from the hyperplane to the closest pattern). 
The corresponding weight vector, normal to the hyperplane, 
has the remarkable property that it can be written as a 
linear combination of some of the training patterns, called 
{\it Support Vectors}. This weight vector ${\bf w}$ minimizes 
the SVM cost function,

\begin{equation}
\label{eq.costprimMSP}
E = \frac{1}{2} {\bf w} \cdot {\bf w},
\end{equation}
subject to the following conditions, imposed to all the patterns 
$\mu=1, \dots,M$, of the training set 
${\mathcal L}_M=\{ ({\bf x}_\mu,x^0_\mu)\}$ with ${\bf x} \in \Re^N, 
x^0_\mu \in \{-1,1\}$,

\begin{equation}
\label{eq.condprimMSP}
x^0_\mu ({\bf w} \cdot {\bf x}_\mu+b) \geq 1,\,\,\,\,   \mu=1,...,M.
\end{equation}
where $|b|$ is the distance of the hyperplane to the origin. Conditions 
(\ref{eq.condprimMSP}) impose that all the patterns be farther than a 
distance $1/\|{\bf w}\|$ from the hyperplane, and minimization of 
(\ref{eq.costprimMSP}) ensures that this distance is maximized.  

It can be shown that the solution to this extremum satisfies 
${\bf w}=\sum_{\mu=1}^M \alpha_{\mu} {\bf x}_{\mu} x_\mu^0$ where 
the coefficients $\alpha_{\mu}$ are nonnegative, and many of them 
are vanishing. If we now introduce a mapping 
${\bf x} \rightarrow \Phi({\bf x})$ the cost function for the SVM 
is given by eq.~\ref{eq.costprimMSP}, but replacing everywhere 
${\bf x}_{\mu}$ by $\Phi({\bf x}_{\mu})$.  

This machines, called hard margin machines, have been succesfully 
analyzed within the approach of statistical mechanics\cite{DOS,BuGo,RGGo1,RGGo2}.

The preceding formulation supposes that the task is linearly separable 
in the working space, as otherwise conditions (\ref{eq.condprimMSP}) 
cannot be fulfilled for all the patterns $\mu$. This may arise either 
because the selected mapping into the feature-space is not adequate, 
or because there is intrinsic noise in the data, and the task cannot 
be learned without training errors. To cope with this problem, a 
modification of the cost function (\ref{eq.costprimMSP}) and the 
conditions (\ref{eq.condprimMSP}) has been suggested~\cite{CoVa}, 
giving  raise to the concept of {\it Soft Margin Classifier}, hereafter 
called SMC.

For simplicity, in this paper we restrict ourselves to 
consider SMCs acting on input space. This is a useful first step 
towards a better theoretical understanding of SMC's with full 
functionality, i.e. using a mapping to a high dimensional feature 
space. We consider classifiers without bias\footnote{With this 
restriction, linearly separable tasks that would need a bias 
become non-realizable. This allows us to explore a larger set 
of non-realizable rules}.

To find the SMC, one has to minimize the function:

\begin{equation}
\label{eq.costprimSMC}
E_{C,k} = \frac{1}{2} {\bf w} \cdot {\bf w}+C \sum_{\mu=1}^M
{\zeta_\mu}^k,
\end{equation}
where $k$ is a positive exponent and $C$ a positive constant, subject 
to the following conditions for $\mu=1,...,M$

\begin{eqnarray}
\label{eq.condprimSMC1}
h_\mu \equiv x^0_\mu \, {\bf w} \cdot {\bf x}_\mu  &\geq& 1 - \zeta_\mu, \\
\label{eq.condprimSMC2}
\zeta_\mu &\geq& 0.
\end{eqnarray}
The {\it slack} variable $\zeta_\mu$ 
is a measure of how much the constraint (\ref{eq.condprimSMC1}) 
is violated for pattern $\mu$. In particular, if $\zeta_\mu>1$ 
then $h_\mu<0$, which means that pattern $\mu$ is wrongly classified. 
Unlike hard margin classifiers, in which all the training patterns are 
excluded from a strip of width $1/\|{\bf w}\|$ on both sides of the separating 
hyperplane, in the case of SMCs, patterns with  $0<\zeta_\mu<1$ lie 
inside this strip, called {\it soft margin}. 

The patterns with $\zeta_\mu>0$, which are the training patterns either 
wrongly classified as well as those correctly classified 
lying within the above mentionned strip, are the Support Vectors.

The exponent $k$ in (\ref{eq.costprimSMC}) is usually set to $1$ or 
$2$ so that the cost function be a quadratic function of the unknowns 
${\bf w}$ and $\zeta_\mu$ ($\mu=1,\dots,M$). Under these conditions, 
the minimum of the cost function is unique~\cite{BuCr}, a fact that gives the SVMs 
a big advantage over other learning algorithms which require a search of 
the lowest of several local minima. 

For practical implementations it is useful to formulate the dual 
problem and use the corresponding Kuhn-Tucker conditions, as 
was done for our simulations. We do not go into further 
details here, as these have been extensively discussed in the 
litterature~\cite{MachineLearn}.

The value of the hyperparameter $C$ in (\ref{eq.costprimSMC}) sets the 
compromise between large margins and small numbers of errors. In practice, 
$C$ should be adjusted, either by trial and error or using more 
sophisticated methods~\cite{Sollich1, Sollich2} to get the best 
performance out of the classifier. 

The paper is organized as follows: in section \ref{sec:typical} 
we clarify what is meant by {\it typical properties} of a classifier 
and give a brief survey of the method used. The learning curves for 
a variety of different tasks are determined and discussed in 
section \ref{sec.results}. We analyze the problems of patterns whose 
classes are random variables (section \ref{sec:random}), patterns classified with 
a rule given by a teacher with the same structure as the trained classifier (section \ref{sec:realizable}), or a different structure 
(section \ref{sec:detunreal}). We also consider tasks where the 
patterns' classes are corrupted by noise (section~\ref{sec:stochastic}). 
We relate the different behaviours of the generalization error 
to simple geometrical properties of the rule to be learned. We also 
present the typical properties of the optimal generalizers, that is, 
obtained with the values of $C$ that minimize the generalization error. 
The main results are summarized in the last section (\ref{sec:conclusions}), 
where we discuss some perspectives of this work.   

\section{What are typical properties ?}
\label{sec:typical} 

Worst case analysis of a learning machine gives exact bounds for different 
quantities of interest, like the generalization error, the training error, the 
number of support vectors, etc. However, very 
often these exact bounds are not tight. In this paper we focus on the 
{\it typical} properties of SMCs faced with particular classes of 
problems. Our results, obtained with the tools of statistical mechanics, 
predict the {\it expected} ({\it averaged} over all the possible training sets) 
behaviour of SMCs. As in the case of perceptrons, this approach allows to get 
insight on the learning properties of the classifiers. The method, thoroughly 
described in a recent book~\cite{EngVdB}, has already been presented 
elswhere in the context of SVMs~\cite{RGGo3}. It was applied for the first time
to learning machines by E. Gardner~\cite{Gardner}, who studied a perceptron 
learning a binary classification task. Statistical Mechanics is generally 
used to determine the properties of the minima of (cost) functions in very 
high dimensional spaces, when the cost depends on a large number of random 
variables, the training patterns. Like in statistical learning theory~\cite{Vapnik}, 
these are assumed to be drawn independently from a probability distribution

Schematically, training a classifier amounts to minimize a cost function 
(which plays the role of an {\it energy}) in the space of the classifier's 
parameters, which in the case of SMCs, are the $N$-dimensional weight vector and 
the $M$ slack variables. This minimization is done using the information contained 
in the set of $M$ training patterns. The typical properties (training error, 
generalization error, etc.) are obtained through averages over all the 
possible training sets corresponding to the task, 
in the limit of very large $N$ and $M$, keeping constant the ratio 
$M/N=\alpha$, hereafter called {\it training set size}. It can be 
proved that in the limit where both $N$ and $M$ diverge (called 
{\it thermodynamic limit}), with $\alpha$ held fixed, these 
averages coincide with the value taken by the considered property  
for {\it almost every training set}. This means that, if we made the 
``experiment" of training a given machine with a given training set 
for given (large enough) $N$ and $M$, we would find that the value of, 
say, the generalization error is very close to the 
one calculated with Gardner's method. Within this context, values of 
$N \simeq 50$ are usually already large. The rigorous validity of the 
techniques used in these calculations has been established 
recently~\cite{Talagrand}. 

\section{Results}
\label{sec.results} 
As already stated, we consider perceptrons trained to minimize the cost 
function (\ref{eq.costprimSMC}) with conditions (\ref{eq.condprimSMC1}) 
and (\ref{eq.condprimSMC2}), with $C>0$. In the following we 
assume, for the sake of simplicity, that the components of the input 
patterns are independently drawn from a gaussian distribution of 
variance $1/\sqrt{N}$:

\begin {equation}
\label{eq.probax}
P({\bf x})=\frac{e^{-N{\bf x}^2/2}}{(2 \pi/N)^{N/2}}.
\end{equation}

We study binary classsification tasks. The label of each pattern, 
denoted by $x^0_\mu \in \{-1,1\}$, is assigned following a rule, which 
may be {\it deterministic} or {\it stochastic}. In the latter case the 
labels are drawn from a probability distribution.  

The training error $\epsilon_t$ is the fraction of patterns in the training set that, 
after training, are incorrectly classified:

\begin{equation}
\label{eq.E_t} \epsilon_t=\frac{1}{M} \sum_{\mu=1}^M
\Theta(-h_\mu).
\end{equation}
where $h_\mu$ is given in equation (\ref{eq.condprimSMC2}).
The generalization error is defined as the probability of misclassifying 
a {\it new} pattern after the network has been trained:

\begin{equation}
\label{eq.E_g} \epsilon_g=\sum_{x^0} \int d{\bf x} P(x^0 | {\bf
x}) P({\bf x}) \Theta(-x^0 \, {\bf w} \cdot {\bf x} )
\end{equation}
where $P(x^0 | {\bf x})$ is the probability  that pattern ${\bf x}$ 
belongs to class $x^0$ (notice that for deterministic rules this is 
a delta function). If the classifier cannot  implement the rule with 
a vanishing generalization error for any training set size $\alpha$ (even
for $\alpha \rightarrow \infty$), the rule is said to be {\it unrealizable}. 
Otherwise, it is {\it realizable}. 

The pertinence of our analytic results has been verified by comparing 
them to numerical simulations. The latter, presented in the 
figures of the following paragraphs, were done for $N=100$ 
and different values of $M$. This value of $N$ is large enough for 
$M/N$ be a good approximation of $\alpha$, which in the theoretical approach 
corresponds to the ratio of $M/N$ in the limit $M \rightarrow \infty$, 
$N \rightarrow \infty$.

We present results for different kinds of rules, starting with the extreme 
case where there is no rule at all, following with the case of a realizable 
rule, and at the end we analyze several examples of non realizable rules. We also 
report results for different exponents $k$ and values of $C$, including 
the optimal value and the limiting case of $C \rightarrow \infty$, 
for each of the rules considered. We call {\it optimal} the value 
$C_{opt}(\alpha)$ that gives, on average, the best generalizer, i.e. 
the lowest generalization error $\epsilon_g$ for each value of $\alpha$. 

\subsection{Patterns with random classes}
\label{sec:random}

In this section we consider that the patterns are labelled randomly. Thus, 
it is impossible for any classifier to predict with any degree of certainty 
the correct label. Therefore, $\epsilon_g \equiv 1/2$. 
Gardner~\cite{Gardner} analyzed with the statistical mechanics approach 
the properties of a perceptron learning such kind of task using the 
number of training errors $M \epsilon_t$ (see eq. (\ref{eq.E_t})) as 
cost function. She obtained the average training error $\epsilon_t$, 
which is the lowest curve in figure \ref{random_task}\footnote{Although 
this curve is not exact, as the approximations used to obtain it break 
down for $\alpha>2$, the corrections to it are believed to be 
small~\cite{ReGy}.}. This curve serves as 
a reference for the performance of the SMC, as by definition it gives 
the lowest training error that can be achieved on average.
It is well known that $\epsilon_t=0$ for $0 \leq \alpha < 2$, which 
means that it is possible to train a perceptron to classifiy correctly 
any set of training patterns only if $\alpha< 2$. The value $\alpha_c =2$ 
is the {\it typical capacity} of the perceptron: for $\alpha > \alpha_c$ 
only a subset of zero measure among all the possible training sets are 
learnable without training errors: the linearly separable ones. But, 
for $\alpha< 2$, a {\it typical} training set has probability $1$ of not being linearly separable.

\begin{figure}
\center{\epsfig{file=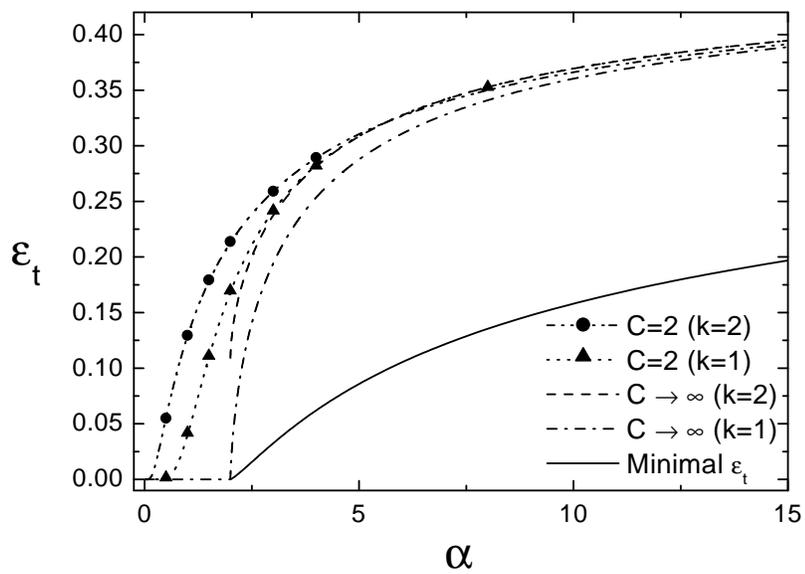,height= 8cm}}
\caption { \small Average value of the training error for different SMCs learning a random task. The points correspond to the result of simulations, averaged over $\sim 100$ different training sets.} 
\label{random_task}
\end{figure}

In figure \ref{random_task} we show our results for the SMCs. For finite 
values of $C$ the average training error does not vanish at any $\alpha$, 
as was to be expected from the fact that SMCs do not aim at minimizing 
this quantity. Moreover, the fraction of training 
errors is rather large compared to the minimal possible values. 
The best performance is obtained in the limit $C \rightarrow \infty$. 
In this limit, the SMC achieves the perceptron's maximal capacity: its 
training error vanishes both with $k=1$ and $k=2$ if $\alpha<\alpha_c$. 
For $\alpha>\alpha_c$ the best performance is obtained using the exponent 
$k=1$ in the cost function (\ref{eq.costprimSMC}). This result is further 
discussed in the conclusions. Notice that when $k=1$ 
the fraction of errors leaves the value $0$ continuously, while the curve 
for $k=2$ presents a discontinuity at $\alpha=\alpha_c$, where it jumps 
from $\epsilon_t=0$ to a finite value $\epsilon_t=0.105$. 

\subsection{A realizable rule}
\label{sec:realizable}

\begin{figure}
\center{\epsfig{file=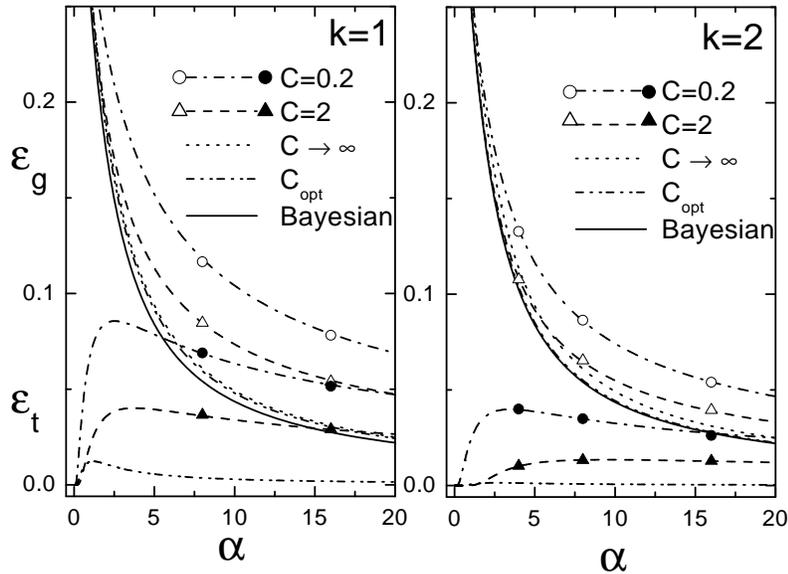,height= 8 cm}}
\caption { \small Average value of the training and generalization errors for different SMCs learning a realizable task. The curve for $C_{opt}$ almost coincides with the one with $C \rightarrow \infty$ for $k=1$ and with the one  of the Bayesian classifier for $k=2$. The points correspond to simulations, averaged over the necessary number of training sets to ensure that the error bars are smaller than the symbols} 
\label{realizable_task}
\end{figure}

Consider now a linearly separable rule so that, at least in 
principle, a perceptron is able to achieve a vanishing generalization 
error. To ensure that the rule is realizable, the labels of the 
examples are given by another perceptron called {\it teacher}:

\begin{equation}
x^0= {\rm sign}\left({\bf w}_0 \cdot {\bf x}\right),
\end{equation}
where ${\bf w}_0$ is the the teacher's weight vector. The 
average training and generalization errors for $k=1$ and $k=2$ are 
represented on figure \ref{realizable_task}, for different values of 
$C$. 

Even though we are considering a realizable rule, which means that it is 
always possible to find a classifier achieving $\epsilon_t=0$, all the 
SMCs (with finite $C$) end up with a finite fraction of training errors.

SMCs with $k=2$ perform better, both in training and in 
generalization, than those with $k=1$. This is so beacuse the term 
proportional to $\zeta_{\mu}^k$ in the cost function (\ref{eq.costprimSMC}) 
penalizes more heavily the errors (which have $\zeta_{\mu}>1$) when $k=2$.

The generalization error of the SMCs has a non monotonic behaviour as a 
function of $C$. On increasing $C$ at any given $\alpha$, $\epsilon_g$ first 
decreases, reaches a minimum value for $C=C_{opt}(\alpha)$, and for 
larger values of $C$ it increases. In the limit $C \rightarrow \infty$, we obtain 
the hard margin solution, which has larger $\epsilon_g$ than the optimum. 

It is interesting to notice that with $C=C_{opt}(\alpha)$, 
which gives (by definition) the smallest 
generalization error, the corresponding training errors are not 
minimal. A similar result has been obtained~\cite{GoGr} for a 
perceptron learning with the algorithm Minimerror~\cite{RafGor}, 
which minimizes a temperature dependent logistic cost function. 
In that case, the parameter that plays the role of $C$ 
is the temperature. It was shown that in the limit of zero temperature 
Minimerror converges to the maximal margin perceptron, which is nothing 
but the hard margin SVM, with $\epsilon_t=0$. However, at {\it finite} 
temperature, the algorithm allows to obtain better generalization 
performance than the hard margin SVM, at the price of making training 
errors. 

For the sake of comparison we included in the same figures the 
generalization error of the bayesian perceptron learning a realizable 
rule~\cite{OpHa}, which is known to be the optimal generalizer. We 
see that for $C=C_{opt}$ the best SMC is obtained with $k=2$. The 
relative difference of its generalization error with respect to 
the bayesian one is at most $1.7\%$ (see fig \ref{delta_eg}). 
Thus, very good generalization is achieved at the expense of some 
training errors. 

In the limit of very large values of $\alpha$, the generalization 
error (that coincides in all cases with that of the training error), 
presents different behaviours. For all finite values of $C$, and both 
for $k=1$ and $k=2$, $\epsilon_g \sim C^{-1/6} \alpha^{-2/3}$. That is, 
$\epsilon_g$ decreases with the training set size {\it slower} than the usual 
$\alpha^{-1}$ law, found in the litterature for zero training error 
solutions. This behaviour changes qualitatively if $C \rightarrow \infty$ 
in which limit we recover the well known~\cite{GoGr} hard margin result 
$\epsilon_g \sim 0.5005/\alpha$. For the curves obtained using $C_{opt}$ 
we obtain: $\epsilon_g(k=1) \sim 0.488 \alpha^{-1}$ and $\epsilon_g(k=2) 
\sim 0.449 \alpha^{-1}$. Despite the fact that $C_{opt}$ is finite, 
the behaviour in this case is proportional to $\alpha^{-1}$ because 
$C_{opt}$ {\it depends on $\alpha$}. 
These results are to be compared with the behaviour of the bayesian 
classifier for large $\alpha$: $\epsilon_g \sim 0.442 \alpha^{-1}$.   

As will be discussed later, the expected value of the first term in the 
cost function (\ref{eq.costprimSMC}) is important to understand the 
behaviour of the learning curves of SMCs. In the present case of a 
realizable rule we obtain, for large $\alpha$, 
$\langle \| {\bf w}\|  \rangle /\sqrt{N} \sim (C \alpha)^{1/3}$.

\begin{figure}
\center{\epsfig{file=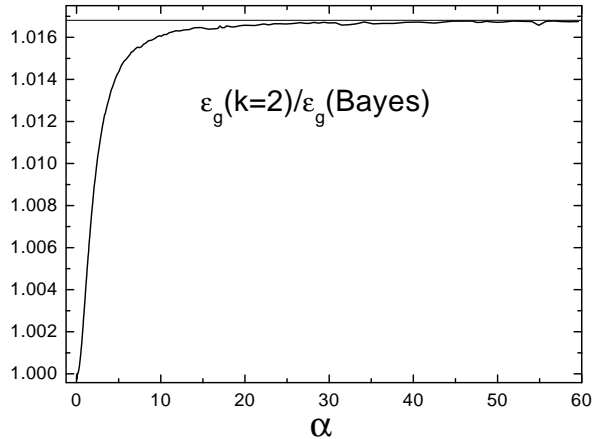,height= 6 cm}}
\caption { \small Comparison between the generalization error for the SMC with $C_{opt}$ and $k=2$ and the generalization error for the bayesian classifier.} 
\label{delta_eg}
\end{figure}

\subsection{Deterministic unrealizable rules}
\label{sec:detunreal} 

\begin{figure}
\centerline{\psfig{figure=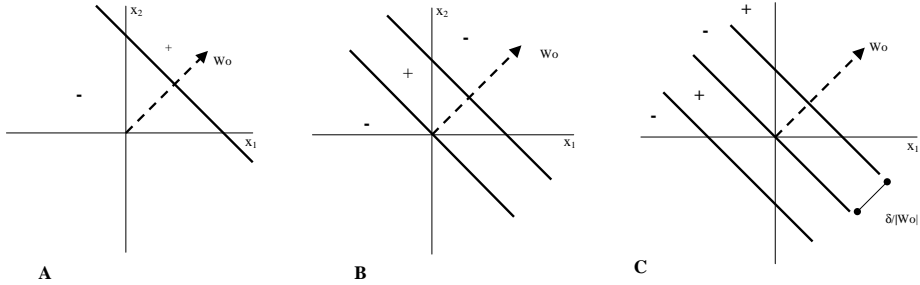,height=4cm}} 
\caption{Three nonlinear rules. A) ${\cal P}(z)=(z-\delta)$ , B)
${\cal P}(z)=z(z-\delta)$, C) ${\cal P}(z)=(z-\delta)z(z+\delta)$}
\label{fig.profs}
\end{figure}

Unrealizable rules are either deterministic, given by ``teachers" 
that have a more complex structure than the ``student", or  
stochastic, which are inherently unrealizable because of 
the randomness involved.

We have studied with great detail the behavior of SMCs facing some
deterministic unrealizable tasks elsewhere~\cite{RGGo4}. Here, we only 
summarize our results. We consider rules corresponding to several 
parallel separating hyperplanes, 
as those sketched on figure \ref{fig.profs}. The class of an 
input vector ${\bf x}$ is given by:  

\begin{equation}
\label{eq.rule} 
x^0 = {\rm sgn}\left({\cal P}({\bf w}_0 \cdot {\bf x})\right).
\end{equation}
where ${\cal P}$ is, in principle, any function of its argument. 
In fact, the rule (\ref{eq.rule}) only depends on the number of zeros 
of ${\cal P}(z)$ and not on its particular expression. We therefore 
assume, without loss of generality, that it is a polynomial. If it 
has $m$ zeros, the rule (\ref{eq.rule}) corresponds to a set of $m$ 
parallel discriminating hyperplanes defined by the equations 
${\bf w_0} \cdot {\bf x}-z_i=0$, where $\{z_i: i=1,...,m \}$ 
are the zeros\footnote{Here, as in the rest of the paper, we call 
zeros the points where the function changes sign.} of ${\cal P}(z)$. 
The distance to the origin of each hyperplane is $|z_i|/ \|{\bf w_0}\|$. 

One quantity of interest is the distribution of stabilities of the 
$M$ training patterns with respect to the SMC solution, defined by 

\begin{equation}
\label{eq.stab} 
\gamma_{\mu} = x^0_{\mu} \frac{{\bf w} \cdot {\bf x_{\mu}}}{\| {\bf w}\| } \equiv \frac{h_\mu}   {\| {\bf w}\|}, \;\;\;\; 1 \leq \mu \leq M.
\end{equation}
If $\gamma_{\mu}>0 (<0)$ the pattern $\mu$ is correctly (incorrectly) 
classified. The norm of $\gamma_{\mu}$ is the distance of pattern 
$\mu$ to the hyperplane orthogonal to ${\bf w}_0$ that contains the 
origin of coordinates. The support vectors  
have $\gamma_{\mu} \leq 1/{\| {\bf w}\| }$. The distribution 
of stabilities of the training patterns, averaged over the 
possible training sets,

\begin{equation}
\label{eq.defrho} 
\rho(\gamma)= \frac{1}{M} {\overline \sum_{\mu=1}^M \delta(\gamma_\mu-\gamma)}.
\end{equation}
gives useful information regarding the SMC's solution. For the 
considered rules, the distribution of stabilities for any finite 
$\alpha$ is nonvanishing everywhere, and has a single discontinuity 
at $\gamma=\sqrt{N}/\| {\bf w} \|$. In addition, if $k=1$ 
there is a Dirac delta at this position, indicating that there is 
a finite fraction of training patterns placed exactly at the SMC's 
margin. The absence of such delta peak for $k=2$ is related to the 
fact that these patterns do not belong to the support vectors in this 
case\footnote{This results from the analysis of the Kuhn Tucker 
equations.}, whereas they are support vectors if $k=1$.
The generalization error and the average fraction of support vectors 
are obtained by performing the integral of $\rho(\gamma)$ between 
$-\infty$ and $0$, and between $-\infty$ and $\sqrt{N}/\| {\bf w} \|$, 
respectively. In fig. \ref{fig.rho(gamma)} we show an example of the 
distribution of stabilities for one particular unrealizable rule.

\begin{figure}
\centerline{\psfig{figure=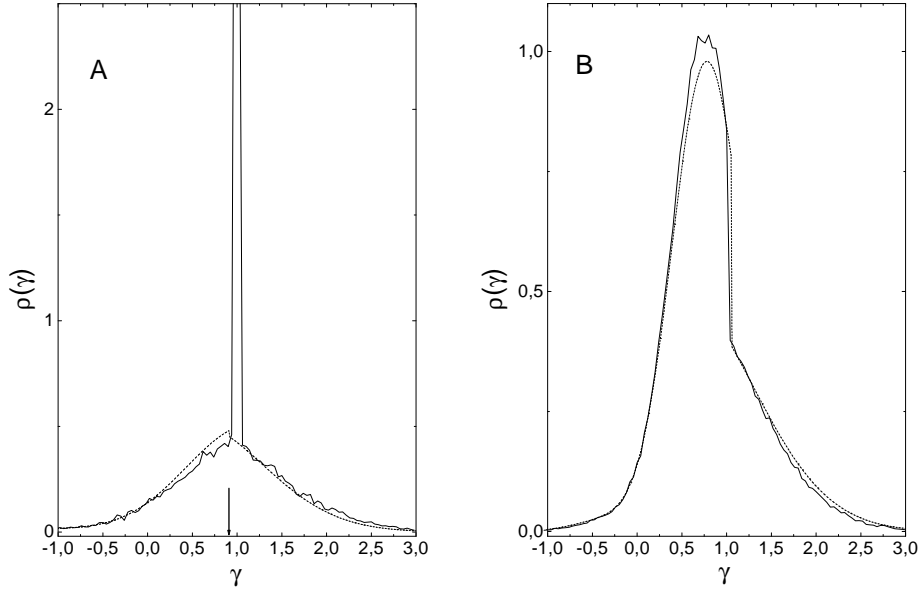,height=8 cm}}
\caption{Distribution of stabilities for two SMCs learning patterns given by the rule with ${\cal P}(z)=z(z-2)$. The full lines represent simulations averaged over 100 training sets, for 200 exemples with $N=100$. The dashed lines are the theoretical predictions for $\alpha=2$.
\newline {\bf A)} $k=1$. The arrow shows the position of the Dirac delta as predicted by the theory.
\newline {\bf B)} $k=2$.} \label{fig.rho(gamma)} 
\end{figure}

We analyzed three types of rules: the {\it linear rule} given by 
${\cal P}(z)=z-\delta$, the ``sandwich" rule ${\cal P}(z)=z(z-\delta)$ 
and the ``reversed-wedge" rule ${\cal P}(z)=(z-\delta)z(z+\delta)$. 
The corresponding learning curves present very different 
behaviours depending on the rules, but, given the rule, they are 
qualitatively similar for both exponents, $k=1$ and $k=2$. 

For the reversed wedge rule with $\delta>\delta_c=\sqrt{2 \ln 2}$ 
and for the sandwich rule, $\epsilon_t(\alpha)$ approaches its finite 
asymptotic value for $\alpha \rightarrow \infty$ from above. 
$\epsilon_g$ decrease rapidly at small $\alpha$, when $\epsilon_t$ is 
still relatively large. In both cases the best generalizer (with 
$C=C_{opt}(\alpha)$) is obtained with the exponent $k=1$. This 
kind of rules are called hereafter rules of type I. 

For the reversed wedge rule with $\delta< \delta_c =\sqrt{2 \log  2}$, 
and the linear rule, hereafter called rules of type II, we find that 
$\epsilon_t$ approaches its asymptotic value from below, whilst the 
decrease of $\epsilon_g$ at small $\alpha$ is slower than for rules 
of type I. The best generalizer (with $C=C_{opt}(\alpha)$) for rules 
of type II is obtained with exponent $k=2$. Figure \ref{fig.dist} 
presents the different kinds of behaviours for some rules of 
type I and II. The behaviour of $C_{opt}$ as a function of $\alpha$ 
is also very different for each type of rules.

\begin{figure}
\centerline{\psfig{figure=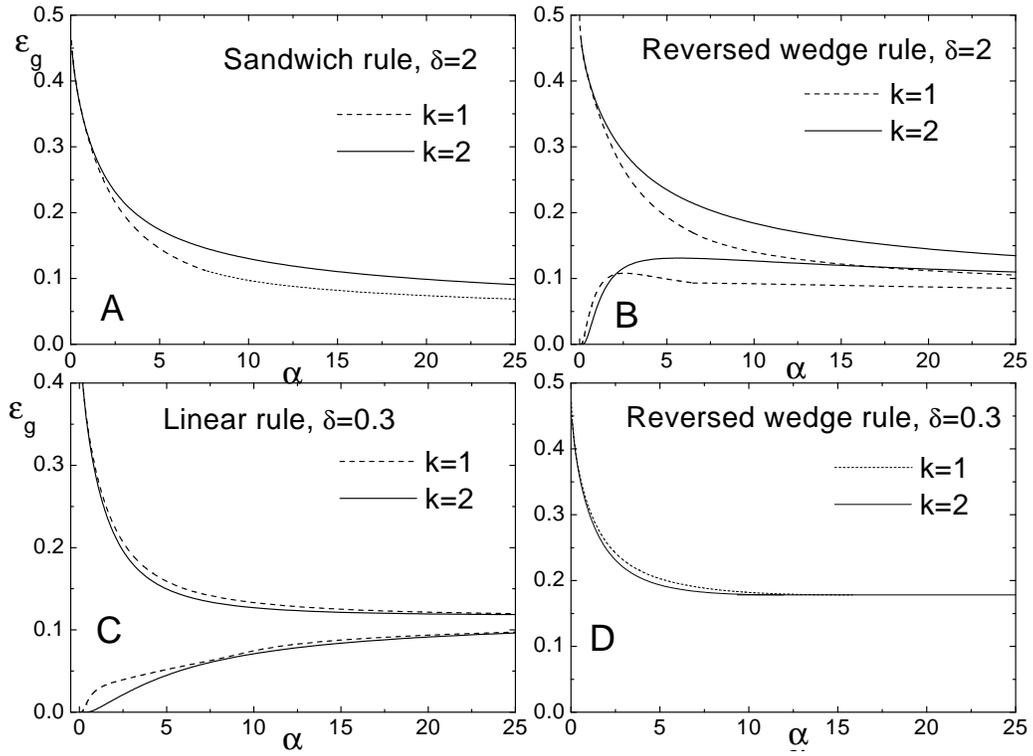,height=10cm}}
\caption{Comparison of the generalization errors of SMCs with $k=1$ and $k=2$ learning unrealizable rules. Figures A and B correspond to rules of type I and figures C and D correspond to rules of type II.} \label{fig.dist}
\end{figure}

In fact, the two types of rules may be characterized by which 
patterns are necessarily misclassified by the ``student" in the 
limit of $\alpha \rightarrow \infty$. In this limit, the weight 
vector of the SMC tends to be aligned either parallel or 
antiparallel to ${\bf w}_0$, the normal to the discriminant 
hyperplanes corresponding to the rule. This is represented on 
figure \ref{fig.type1y2}, where the misclassified patterns lie 
in the shaded regions. For rules of type I, these regions are 
{\it unbounded} half-spaces. On the other hand, for type II 
rules errors are restricted to the {\it bounded} regions close 
to the origin. This remark allows to understand why the best 
performances in generalization are obtained with different 
exponents $k$ depending on the type of rule. In 
general, the student's (unique) hyperplane is rotated with respect to the 
teacher's vector ${\bf w}_0$ by an angle that depends on the type of 
rule and on the exponent $k$. If the training errors lie in the unbounded 
regions, the rotation angle with $k=2$ 
will be larger than with $k=1$ because this reduces the cost of the 
errors located far from the hyperplane. If this kind of errors cannot 
be avoided, as arises in rules of type I, the generalization error with $k=2$ 
will be larger than with $k=1$. On the other hand, when the unavoidable 
errors are relatively close to the origin of coordinates, their larger 
cost when using $k=2$ will induce an orientation of the hyperplane closer 
to ${\bf w}_0$ than with $k=1$. This results in a better generalization 
performance with $k=2$ than with $k=1$.

\begin{figure}
\centerline{\psfig{figure=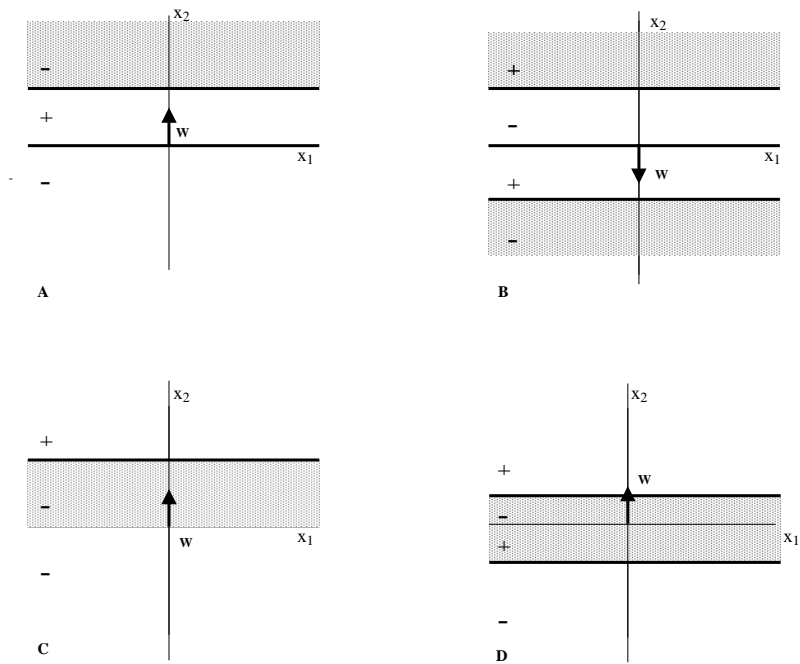,height=10cm}}
\caption{Comparison of the regions with errors for $\alpha \rightarrow \infty$, of SMCs learning:
\newline {\bf A)} Sandwich rule (type I),
\newline {\bf B)} Reversed wedge rule, with $\delta>\delta_c$ (type I),
\newline {\bf C)} Linear rule (type II),
\newline {\bf D)} Reversed wedge rule, with $\delta<\delta_c$ (type II).
\newline The arrows indicate the asymptotic orientation of the hyperplane of the SMC. In the shaded region the patterns are incorrectly classified.} \label{fig.type1y2}
\end{figure}

To conclude this study of deterministic unrealizable rules, we discuss some 
general results. In particular, the different rules can be characterized by 
a single quantity:

\begin{equation}
\label{eq.avstab} \langle \gamma_0 \rangle = \int d\gamma \,
\rho(\gamma) = \int Dz \,z \,{\rm sign}({\cal P}(z))= \sum_{i=1}^m
(-1)^i e^{-z_i^2/2},
\end{equation}
where the $z_i$ are the zeros of ${\cal P}(z)$. $\langle \gamma_0 \rangle$ 
represents the average stability of the training patterns with 
respect to a hyperplane normal to ${\bf w}_0$ passing through the origin. 

One interesting result is that if the rule is such that 
$\langle \gamma_0 \rangle =0$ then the SMC (for all values 
of $C$ and $k$) is unable to generalize: $\epsilon_g=1/2$ 
for all values of $\alpha$, even though the training error 
remains small. This phenomenon is known as {\it memorization 
without generalization}~\cite{HaMaMe}. A similar result has 
been obtained by Reimann and van den Broeck~\cite{REVdB} when 
the classifier uses Hebb's rule. 

\begin{figure}
\centerline{\psfig{figure=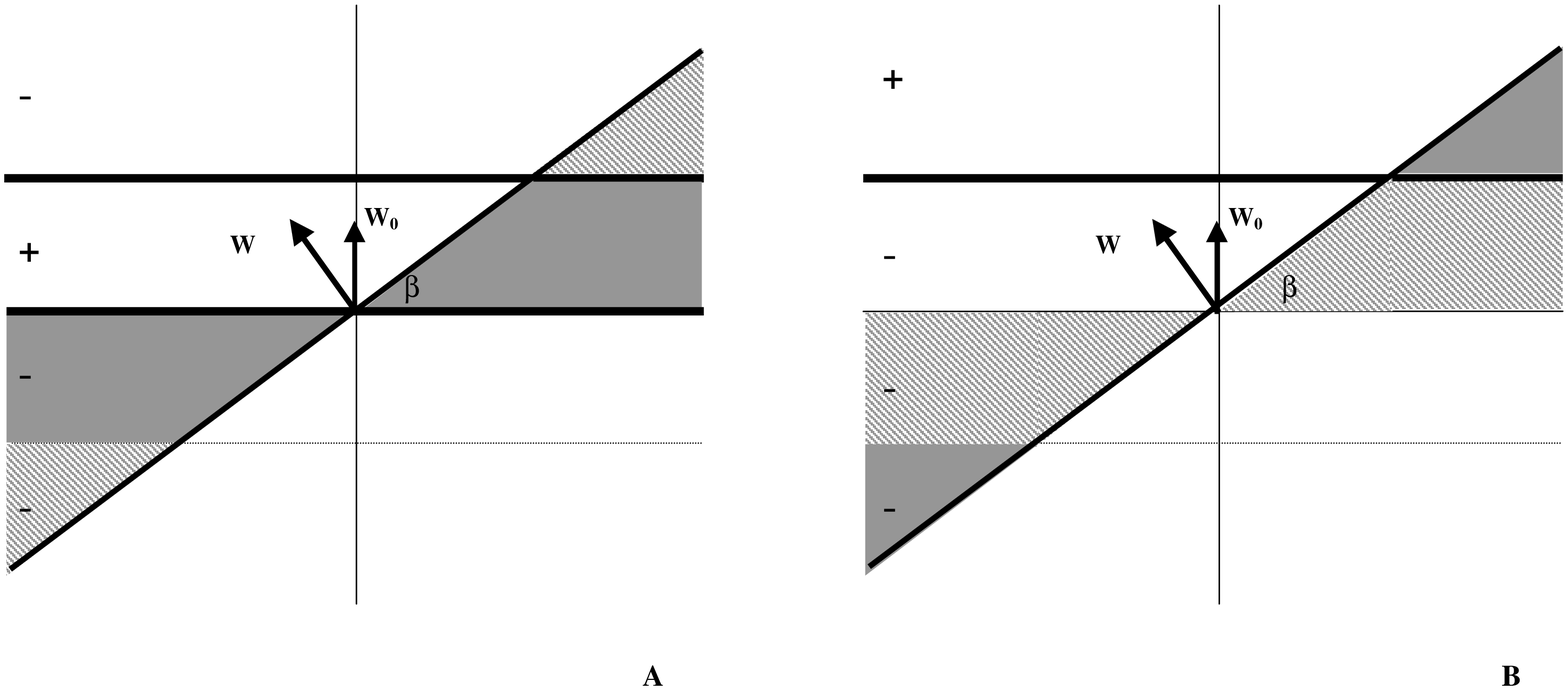,height=9 cm}}
\caption{Comparison of two rules. The thick horizontal lines represent the hyperplanes of the teacher. The signs + and - are the classes assigned by the teacher inside the horizontal bands shown. The line perpendicular to ${\bf w}$ represents the SMC's hyperplane. For $\alpha \gg 1$, the angle $\beta$ between ${\bf w}$ and ${\bf w}_0$ is $\ll 1$. The shaded regions contain the misclassified examples. 
\newline {\bf A)} Rule with ${\cal P}(z)=z(z-\delta)$.
\newline {\bf B)} Rule with ${\cal P}(z)=z-\delta$.}
\label{fig.sombra}
\end{figure}

For large values of $\alpha$ the weight vector of the SMC tends to align 
parallel to ${\bf w}_0$ if $\langle \gamma_0 \rangle >0$ and antiparallel 
if $\langle \gamma_0 \rangle <0$. Notice that this does not imply that 
the best generalization performance is reached asymptotically for 
$\alpha \rightarrow \infty$, as it can be shown that $\epsilon_g$ is not 
a monotonic function of $\alpha$. The asymptotic value of the generalization 
error is

\begin{equation}
\label{eq.Easy} \epsilon_g^{\infty}=\epsilon_g(\pm1)=\int Dz
\Theta(\mp z \, {\cal P}(z))
\end{equation}
which can be even larger than $1/2$. This means that using the SMC for 
large $\alpha$ can be worse than classifying the patterns randomly. The 
asymptotic behaviour of $\epsilon_g$ only depends on whether 
${\cal P}(0)=0$ or ${\cal P}(0) \neq 0$, that is, whether 
one of the rule's hyperplanes contains the origin or not. 
If ${\cal P}(0)=0$, we obtain a power law: 
$\epsilon_g-\epsilon_g^{\infty} \propto \alpha^{-1/2}$. The same exponent 
was found by Fontanari and Meir~\cite{MeFo2} for a machine learning 
a realizable rule with an algorithm accepting training errors. 
If ${\cal P}(0) \neq 0$ the convergence is exponential: 
$\epsilon_g-\epsilon_g^{\infty}  \sim \alpha^3 e^{-z_0^2 \alpha}$, 
where $z_0$ is the zero of ${\cal P}(z)$ with smallest norm. 
The existence of these two regimes is related to whether the patterns 
contributing to $\epsilon_g$ lie close to or far from the student's 
hyperplane. Figure \ref{fig.sombra} presents examples of both regimes. 
The angle $\beta \ll 1$ shown in the figure gives the
orientation of the SMC's hyperplane relative to that of the teacher, 
for large $\alpha$. Within our approach, this angle can be calculated 
as function of $\alpha$. In fig. \ref{fig.sombra}A, we show the 
situation corresponding to the rule with ${\cal P}(z)=z(z-\delta)$. 
Consider the difference $\epsilon_g({\bf w}_0)-\epsilon_g({\bf w})$ 
of the generalization error with respect to its asymptotic value. It 
is proportional to the fraction of patterns that in the dark 
grey regions because the contributions of the light grey regions 
compensate each other. As the patterns' distribution is gaussian, 
the main contribution is due to the fraction of points close to 
the origin, which is roughly proportional to the angle $\beta \ll 1$. 
In fig. \ref{fig.sombra}B, we show the rule with ${\cal P}(z)=z-\delta$. 
Here again, the difference of generalization error is given by the points 
inside the dark grey regions, that in this case are placed far from 
the origin (the contributions from the light grey regions compensating 
each other). The fraction of points in this region is roughly 
proportional to $\sim \exp(-(\delta/\beta)^2/2)$. 

The constants involved in the asymptotic terms do not depend on $C$. 
This can be understood from the fact that in the limit of large 
$\alpha$, the complexity term ${\bf w}^2/2N$ in the cost function 
tends to a constant and therefore it is the error term that dominates 
completely, thus turning $C$ into a multiplicative constant to the cost. 
Notice that this is not the case for the realizable rule, where we 
have that ${\bf w}^2/N \rightarrow \infty$ if $\alpha \rightarrow \infty$.

We have also calculated the behavior of the quantities of interest in 
the limit of very small number of examples, $\alpha \ll 1$. We find 
that the norm of the weight vector increases with $\alpha$ like 
$\| {\bf w} \| /\sqrt{N} \sim C^2 \alpha$. The generalization error 
decreases as $1/2-\epsilon_g \sim \langle \gamma \rangle^2 \alpha^{\frac{1}{2}}$. 
This interesting result shows that for a small number of examples the 
SMC has some generalization power even for those rules where the asymptotic 
value of $\epsilon_g$ for large $\alpha$ is worse than if the new inputs 
were randomly classified.

Another interesting result is that for the SMC with $k=1$, the fraction of 
support vectors that lie exactly on the margin tends to $1$ for 
$\alpha \rightarrow 0$ if $C>1$, but to $0$ if $C<1$.

\subsection{Stochastic rules}
\label{sec:stochastic}

In this section we consider rules that do not determine univocally 
the class of the patterns. In particular, we study rules where the 
output of the teacher is corrupted by a random {\it noise}. When 
the noise is {\it additive} the class of pattern ${\bf x}$ is

\begin{equation}
\label{eq.gausnoi} x^0_+ = {\rm sgn}\left({\cal P}({\bf w}_0 \cdot
{\bf x} + \eta)\right),
\end{equation}

\noindent where $\eta$ is a random variable drawn from a distribution 
which we assume is a {\it gaussian}  of variance $\Delta$. If the 
noise is {\it multiplicative}, the class given is

\begin{equation}
\label{eq.multnoi} x^0_{\times} = {\rm sgn}\left({\cal P}({\bf
w}_0 \cdot {\bf x} \, \eta)\right),
\end{equation}

\noindent where $\eta = \pm 1$ with $P(\eta=1)=p$ and $P(\eta=-1)=1-p$. 
The effect of these two types of noise is the same as that of a 
corruption of the pattern to be classified. The gaussian noise 
only changes the class of patterns that are close to the separating 
hyperplanes whereas with the multiplicative noise the probability 
of changing the class of a pattern with respect to the deterministic 
rule does not depend on distances. The effect of additive noise has 
been studied using the statistical mechanics approach in the case of 
a perceptron learning a rule with ${\cal P}(z)=z$ by Gyorgyi and 
Tishby~\cite{GyTi} and by Opper and Haussler~\cite{OpHa} in their analysis 
of the bayesian perceptron.

The asymptotic behaviors we obtain for the SMCs learning noisy rules 
are qualitatively the same as the ones we obtained for unrealizable 
rules. The relative norm of the classifier, ${\bf w}/\sqrt N$, tends 
to a constant value when $\alpha \rightarrow \infty$ for both types 
of noise. The alignement of the weight vector with the vector of the 
rule depends on the sign of $\langle \gamma_0 \rangle$. 

The generalization error shows different asymptotic behaviours, depending 
on the type of noise. For multiplicative noise we obtain a power law 
behaviour if one of the hyperplanes of the rule contains the origin 
of coordinates and an exponential decay otherwise, like in the case 
of unrealizable rules. On the other hand, for gaussian additive noise, the 
rate of convergence does not depend on ${\cal P}(0)$ and is 
a power law for all the rules: 
$\epsilon_g-\epsilon_g^{\infty} \sim \frac{\alpha^{-1}}{\psi} 
\sum_{i=1}^m (-1)^i \exp(-\frac{x_i^2}{2 \psi})$ 
where $m$ is the number of zeros of ${\cal P}(z)$ and 
$\psi=\sqrt{1+\Delta^2}/\Delta$. The suppression of the 
exponential convergence that exists in the case of a 
deterministic unrealizable rule with no teacher's hyperplane 
passing thorugh the origin can be understood in the same way 
as before. The important point is that additive noise 
alters the class of patterns that are close to the 
hyperplane. This introduces errors inside the central 
strip which are enough to suppress the exponential convergence, 
which is recovered in the limit of vanishing noise, 
when $\Delta \rightarrow 0$. Notice that the multiplicative 
noise does not introduce any errors inside the central band. 

The learning curves for the particular case of rules with 
${\cal P}(z)=z$ with multiplicative noise are shown in 
figure \ref{fig.mult}. Figure \ref{fig.gaus} presents results 
for gaussian additive noise. We can see for these rules the 
same effects we observed for deterministic rules of type I 
and II respectively. Geometrically, the reason is clear: as 
the gaussian noise changes mostly the class of the patterns 
that are near the hyperplanes of the teacher, the errors made 
in this case will be bounded, whereas for multiplicative noise 
the alteration of the class does not depend on the distance, producing 
unbounded errors.

\begin{figure}
\centerline{\psfig{figure=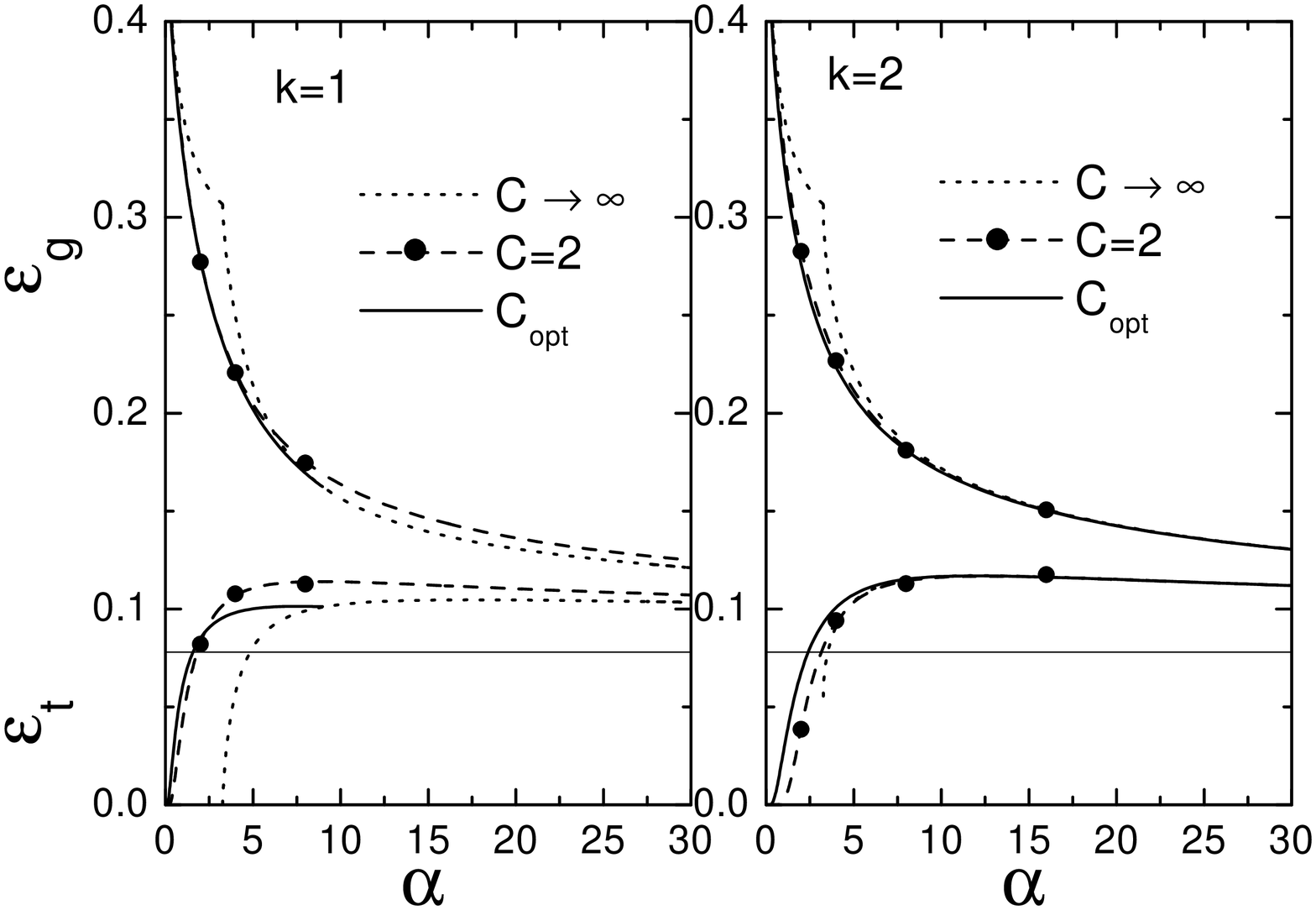,height=8 cm}} \caption{Training and generalization errors for an SMC learning a rule with ${\cal P}(z)=z$ corrupted by multiplicative noise ($P(\eta=1)=0.922$). The horizontal line shows the asymptotic value of both errors. The symbols represent simulations made with N=100, averaged over enough training sets to make the error bars smaller than the symbols.}
\label{fig.mult}
\end{figure}

\begin{figure}
\centerline{\psfig{figure=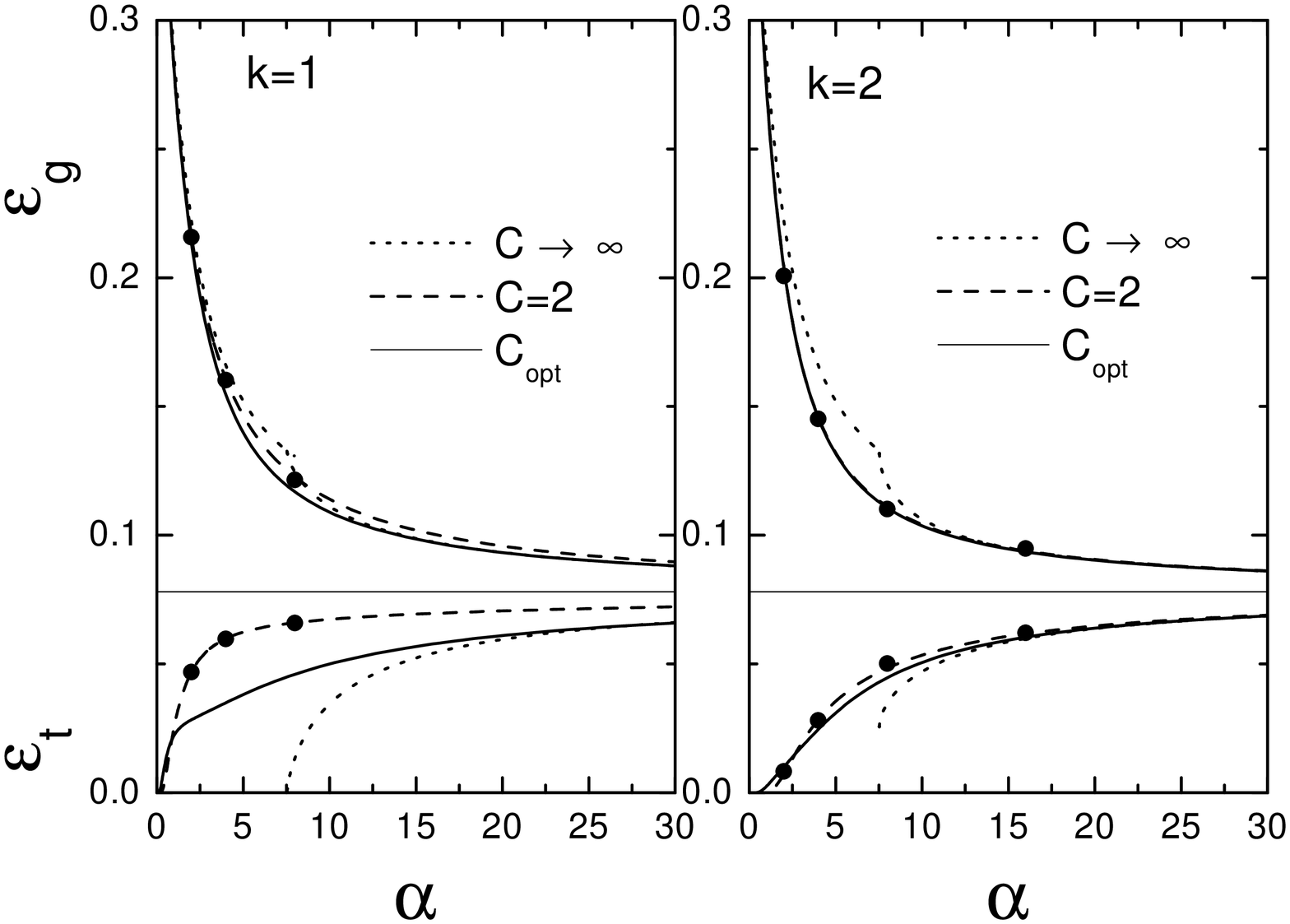,height=8 cm}} \caption{Training and generalization errors for an SMC learning a rule with ${\cal P}(z)=z$ corrupted by additive gaussian noise ($\eta=0.97$). The horizontal line shows the asymptotic value of both errors. The symbols represent simulations made with N=100, averaged over enough training sets to make the error bars smaller than the symbols.}
\label{fig.gaus}
\end{figure}

\section{Conclusions}
\label{sec:conclusions}
We have studied the typical properties of the Soft Margin Classifiers, using 
the tools of Statistichal Mechanics, for several different scenarios. This 
approach allowed us to study also the properties of the optimal classifier, 
which is the one obtained when the hyperparameter $C$ is tuned to obtain 
the lowest value of the generalization error. It turns out that, for 
realizable rules, the classifier obtained with $C_{opt}$ is very close 
to the optimal performance, given by Bayesian learning. The best results 
are obtained with an exponent $k=2$ for the slacks in the cost function; 
the relative difference between both learning curves is smaller 
than $1.7\%$, for all $\alpha$. 

As the generalization error cannot be known exactly in practice, 
the values of $C_{opt}$ that we have obtained are only useful 
to provide a reference generalization error curve, against which 
the performance of the various algorithms devised to optimize $C$ 
may be tested. 

In general, when the rule is non realizable and the SMC cannot 
avoid misclassification of some training patterns, the learning 
curves present two types of behaviours. In rules of type I, in 
which these unavoidable errors lie at large distances of the 
SMC's hyperplane, the generalization performance is better 
if $k=1$ is used as exponent of the slack variables in the 
cost function. Conversely, if errors are confined to a strip 
containing the origin, like in rules of type II, it may be 
more convenient to use a SMC with $k=2$. This is due to the 
way errors are weighted in the cost function, and may be used as 
a rule of the thumb for applications, as it only uses as a criterion 
the distances of the misclassified patterns to the discriminating 
surface. 

In the case of an SMC learning patterns whose classes are entirely 
random, the best learning performance is achieved for $k=1$. This 
is not surprising, as this case is similar to the rules of type I, 
because as the classes are random, the errors made by the classifier 
are evidently unbounded.  

For the unrealizable rules considered, the convergence of the 
training and generalization errors to their asymptotic values 
as a function of $\alpha$ in the limit $\alpha \gg 1$, 
follows either an exponential or a power law decay with 
exponent $1/2$, depending on whether or not one of the teacher's 
hyperplanes contains the origin. If there is a gaussian additive 
noise, only the power law decay exists. It would be interesting 
to determine if this two types of asymptotic behaviours are universal 
for all the unrealizable rules or if there are still more possible 
regimes. It is remarkable that even though the asymptotic value of 
$\epsilon_t$ and $\epsilon_g$ can be larger than one half (which 
is worse than that achieved by randomly classifying the patterns), 
in the regime of small values of $\alpha$ we have always 
$\epsilon_t<1/2$ and $\epsilon_g<1/2$. 

The statistical mechanics approach can be extended to consider more 
complicated (and probably more realistic) pattern distributions, like 
biased or non-gaussian distributions. The bias $b$ can be included, but 
the calculations become much more complicated. If a bias is allowed, 
probably the asymptotic exponential behavior mentioned above would 
disappear, because in this case the hyperplane of the classifier can 
be shifted until it coincides with one of the hyperplanes of the 
considered rule. Classifiers using $k=3$ in the cost function can also 
be studied within this approach, in much the same way as $k=1$ and $k=2$, 
and they may present some interesting features, even though they 
are not used in practice. The model can be extended to include the 
mappings from the input space to a feature space, but at the expense 
of considerably increasing the complexity of the calculations.

\subsection*{Acknowledgments}
We acknowledge discussions with the participants to the workshop 
``Statistical Mechanics of Information Processing in Cooperative 
Systems", held in the Max Planck Institute f\"ur Komplexer Systeme 
in Dresden, on March 5-23, 2001.

\end{document}